\documentclass[12pt,preprint]{aastex}
\slugcomment{2.\ January 2007, draft for ApJ}

\lefthead{Kamp, Freudling \& Chengalur}
\righthead{H\,{\sc i} in protoplanetary/debris disks}

\received{2006 August 7}
\begin{document}
\title{H\,{\sc i}~21~cm emission as a tracer of gas during the evolution from protoplanetary to debris disks}
\author{I.~Kamp}
\affil{Space Telescope Science Division of ESA, STScI, 3700 San Martin Drive,
Baltimore, MD 21218, USA, e-mail: kamp@stsci.edu}
\author{W. Freudling}
\affil{ST-ECF, European Southern Observatory, Karl-Schwarzschild-Str. 2, D-85748 Garching bei M\"unchen, Germany}
\author{Jayaram N Chengalur}
\affil{ATNF/CSIRO, P. O. Box 76, Epping NSW 1710, Australia\\
NCRA (TIFR), Pune University Campus, Postbag 3, Ganeshkhind, Pune 411007 India}

\begin{abstract}
We present models for the HI 21~cm emission from circumstellar disks and use them to convert observed upper limits on the H\,{\sc i}~21~cm flux to limits on the total disk gas mass. The upper limits we use come from earlier Australia Telescope Compact Array observations of the debris disk around $\beta$~Pictoris as well as fresh Giant Meterwave Radio Telescope observations of HD\,135344, LkCal15 and HD\,163296. Our observations and models span a range of disk types,
from young proto-planetary disks to old debris disks. The models self-consistently calculate the gas chemistry (H/H$_2$ balance) and the thermal structure of UV irradiated disks.  Atomic hydrogen production is dominated by UV irradiation in transition phase objects as well as debris disks, but for very young disks, HI production by stellar X-rays (which we do not account for) is important. We use a simple radiative transfer approach to convert the model disk parameters into predicted H\,{\sc i}~21~cm line maps and spectral profiles. This allows a direct comparison of the observations to the model. We find that the H\,{\sc i} traces the disk surface layers, and that the H\,{\sc i} emission, if detected, could be used to study the effects of irradiation and evaporation, in addition to the kinematics of the disk. Our models cover massive protoplanetary disks, transition phase disks and dusty debris disks. In massive protoplanetary disks, UV produced H\,{\sc i} constitutes less than 0.5\% of the total disk mass, while X-rays clearly dominate the chemistry and thus the H\,{\sc i} production. For the two such disks that we have observed, viz. those around LkCa\,15 and HD\,163296, the predicted 21~cm flux is below the current detection limit.  On the other hand, transition phase disks at distances of 100~pc have predicted 21~cm fluxes that are close to the detection limit. Finally, in debris disks, hydrogen is mainly molecular since the high dust-to-gas mass ratio leads to warmer disks, thus increasing the formation rate of H$_2$. This, in conjunction with the small total gas mass, makes the predicted flux to fall below our detection limit. However, future telescopes like the SKA should be able to image the H\,{\sc i}~21~cm emission from nearby transition phase disks, while a radio telescope with only $\sim 10\%$ the area of the SKA should be able to detect the emission from such disks. Such 21~cm line observations would probe both disk evaporation as well as disk kinematics.
\end{abstract}


\section{Introduction}

Protoplanetary disk evolution and planet formation are closely intertwined. To understand planet formation, we first need to develop a comprehensive picture of the physical and chemical conditions in  protoplanetary disks. While dust has been observed over the entire range of disk evolution --- from the very young massive disks out to the old debris disks that contain only a few lunar masses of small dust grains ---,  gas is much more difficult to detect. Hence our picture of the chemical evolution of the gas phase is very incomplete and mainly restricted to a few prototypes. This is unfortunate, because gas plays an important role in the disk physics. It determines the hydrostatic equilibrium in the early phases, it affects the dust grain growth and dynamics as well as planet migration, and eventually its dispersal ends the planet formation process.

Hydrogen is the most abundant element in molecular clouds and in the protoplanetary disks that are a byproduct of the star formation process. Most of the disk mass is in the form of either atomic or molecular hydrogen. Depending on the number of photons capable of ionizing hydrogen (i.e. photons with energies larger than 13.6~eV), there will be a low density layer of ionized hydrogen at the surface of the disks. In the following, we ignore this layer as its contribution to the disk mass is negligible. Underneath this H\,{\sc ii} layer, there will be a layer dominated by far ultraviolet photons from the star where hydrogen is mostly neutral. As the optical depth becomes large enough to shield this radiation, molecular hydrogen dominates. Even though molecular hydrogen dominates the disk mass in early phases of disk evolution, it is difficult to detect; due to the lack of dipole transitions, its ro-vibrational lines are very weak and the line-to-continuum ratio in the near-IR limits the sensitivity. Atomic hydrogen has electronic transitions, which need typical excitation temperatures in excess of 5000~K and therefore originate only in the surface layers very close to the central star, where the gas can be that hot. In this paper we study the role of the H\,{\sc i}~21~cm line --- a hyperfine transition of atomic hydrogen --- for detecting the gas in various evolutionary stages of protoplanetary disks.

The H\,{\sc i} line could be an important tool to study the surface layers of young protoplanetary disks. By simultaneous observations of the H\,{\sc i}~21~cm line and ro-vibrational lines of molecular hydrogen, we can observationally determine the H\,{\sc i}/H$_2$ fraction and compare it to predictions from disk models. This constrains both the photochemistry in the disk surface and the strength of the dissociating stellar (and surrounding)  radiation field.  The H\,{\sc i} line is also a very important tool in tracing disk dispersal and in particular photoevaporation processes. \citet{Hollenbach:2006} have proposed that the FUV photoevaporation from the star itself is one of the main processes of disk dispersal. The FUV radiation can penetrate much deeper into the disk than the EUV radiation and it heats the gas up to temperatures of a few 100 K. \citet{Adams:2004} have shown that photoevaporation can already start at temperatures much lower than the virial temperature, $T_{\rm gas} \sim 0.2 --- 0.5\,T_{\rm crit}$. The critical temperature is given by
\begin{equation}
T_{\rm crit} = G\,M_\ast m_{\rm H}/(k\,r)\,\,\,{\rm K},
\end{equation}
where G is the gravitational constant, $M_\ast$ the stellar mass, $m_{\rm H}$ the mass of an hydrogen atom, $k$ the Boltzmann constant and $r$ the radial distance from the star. Typical values for $T_{\rm crit}$ are in the range of 180 to 10\,000~K for our T Tauri disk models. This means, that the FUV radiation field can drive a photoevaporative wind from layers much deeper in the disk than the EUV and hence this wind would be more massive than the EUV wind and dominate the mass loss of the disk. H\,{\sc i} observations could thus be used to trace this mass loss and to derive photoevaporation rates, which can be compared to the theory.

The rest of this paper is organised as follows. The Australia Telescope National Facility (ATNF) and the Giant Meterwave Radio Telescope (GMRT) H\,{\sc i}~21~cm line observations are described in Sect.~\ref{observations}. In Sect.~\ref{hfrac} we derive approximate estimates of the fraction of atomic hydrogen in circumstellar disks at different stages of evolution. Detailed models of each evolutionary stage ranging from protoplanetary disks to debris disks are presented in  Sect.~\ref{disk models}  while in Sect.~\ref{radtrans}  we outline the details of the H\,{\sc i}~21~cm line radiative transfer. In
Sect~\ref{predictions} we compare the model predictions with the observations, and finally in Sect.~\ref{discussion} we assess the prospects of using future radio telescopes to observe H\,{\sc i}~21~cm line emission to trace gas in circumstellar disks at various evolutionary stages.

\section{Observations}
\label{observations}

Atomic hydrogen 21~cm line emission has never been detected in circumstellar disks. The only case for which upper limits are available in the literature is $\beta$~Pictoris, which has been observed at ATCA by \citet[F95]{Freudling:1995}. This program was set up as a pilot survey. Our sample was hence selected to include  disks from which HI was likely to be detectable (based on the presence of detectable amounts of molecular gas), while at the same time trying to cover a wide range of stellar and disk types.

We have carried out 21~cm line observations of the disks around  HD\,135344, LkCa15 and  HD\,163296 at the {\em Giant Metrewave Radio Telescope} operated by the National Centre for Radio Astrophysics, India. The stellar and disk classification and properties are summarized in Table~\ref{sources}. The observations were conducted on August 29 and 30, 2004. The total integration time on each source was 4 hours. We used a 1~MHz bandpass with 128 channels, which results in a channel width of
1.6~km/sec.  The standard calibrators 3C 232 and 3C 286 were observed at the start and end of the observing run and used to calibrate the visibility amplitudes and the bandpass shape. Phase calibration was carried out with continuum sources 1522-275, 0431+206 and 1751-253 for  HD\,135344, LkCa15 and HD\,163296, respectively. For each disk, separate maps were produced at angular resolutions of 2"$\times$3" and 7"$\times$9" using appropriate UV ranges and tapers. The spectra at the position of the disks  were in each case closely inspected for any signs of H\,{\sc i} emission. Finally, all spectra were also smoothed with a 3 channel boxcar function. No evidence of any line emission was found in either the smoothed or unsmoothed spectra. One $\sigma$ upper limits are listed in Table~\ref{upperlimits}.

\clearpage
\begin{table}[htdp]
\caption{Observed stellar and disk parameters: $M_{\rm disk}$(1.3mm) and $M_{\rm disk}$(CO) are the
total disk masses as derived from 1.3~mm continuum and CO radio observations respectively.}
\begin{center}
\begin{tabular}{llllllllll}
star      & classification & d     & i                &  T$_{\rm eff}$ &    R$_\ast$      &   M$_\ast$       & M$_{\rm disk(1.3mm)}$ & M$_{\rm disk(CO)}$  & L$_{\rm X}$\\
            &          & [pc]  & [$^\circ$] &    [K]                 &    [R$_\odot$] &   [M$_\odot$]  &  $10^{-2}$~M$_\odot$    & $10^{-2}$~M$_\odot$ & [erg s$^{-1}$] \\[2mm]
\hline
$\beta$~Pictoris  & A5V debris disk & 19.3$^1$ & 0$^6$ & 8200$^1$ & 1.47$^1$ & 1.75$^1$ & 0.003$^4$ & \ldots & $2.6\times 10^{25}$~$^7$\\
HD163296    & A1Ve pre-MS & 122$^2$ & 65$^3$ & 9230$^2$ & 2.7$^2$ & 2.4$^3$ & 6.5$^4$ & 0.056$^4$ & $4 \times 10^{29}$~$^8$\\
HD135344    & F4Ve pre-MS & 84$^2$ & 60$^3$ & 6200$^2$ & 1.3$^2$ & 1.3$^3$ & 0.28$^4$ & 0.00021$^4$ & \ldots\\
LkCa15      & K5 pre-MS & 140$^4$ & 60$^5$ & 3980$^4$ &  &  & 3.3$^4$ & 0.014$^4$ & $<4 \times 10^{29}$~$^9$\\
\end{tabular}
\end{center}
\tablerefs{ \\
$^1$ \citet{Crifo:1997};  $^2$ \citet{Jaya:2001}; $^3$ \citet{Dominik:2003}; $^4$ \citet{Thi:2001}; $^5$ \citet{Qi:2003}; $^6$\citet{Smith:1984}; $^7$\citet{Hempel:2005}; $^8$\citet{Neuhaeuser:1995} ; $^9$\citet{Swartz:2005}}
\label{sources}
\end{table}

\begin{table}[htdp]
\caption{Observed 21 cm line emission upper limits}
\begin{center}
\begin{tabular}{cccrc}
star     & beamsize      &  $\Delta$v    &  $1\sigma$ limit & source \\
            & [" $\times$ "] & [km/sec]       &   [mJy/beam] &        \\[2mm]
\hline
$\beta$~Pictoris  & 27$\times$ 25  &   1.7        &  3.7             & F95     \\
            & 27$\times$ 25  &   8.3        &  1.7             & F95     \\
            & 68$\times$ 65  &   1.7        &  10.8            & F95     \\
            & 68$\times$ 65  &   8.3        &  4.7             & F95     \\
HD163296    & 3 $\times$ 2   &   1.6        &  3.2             &this work \\
            & 9 $\times$ 7   &   1.6        &  13.7            &this work \\
            & 3 $\times$ 2   &   4.8        &  2.0             &this work \\
            & 9 $\times$ 7   &   4.8        &  10.1            &this work \\
HD135344    & 3 $\times$ 2   &   1.6        &  4.6             &this work \\
            & 9 $\times$ 7   &   1.6        &  12.5            &this work \\
            & 3 $\times$ 2   &   4.8        &  2.6             &this work \\
            & 9 $\times$ 7   &   4.8        &  10.6            &this work \\
LkCa15       & 3 $\times$ 2   &   1.6        &  7.6             &this work \\
            & 9 $\times$ 7   &   1.6        &  20.0            &this work \\
            & 3 $\times$ 2   &   4.8        &  5.1             &this work \\
            & 9 $\times$ 7   &   4.8        &  11.6            &this work \\
\end{tabular}
\end{center}
\label{upperlimits}
\end{table}
\clearpage

\section{Estimates of the H\,{\sc i} mass fraction}
\label{hfrac}

In the following we do not attempt to generate disk models that fit all the observed stellar and disk properties (see Table~\ref{sources}), but rather use generic representative disk models for T Tauri disks, Herbig Ae disks, transition phase and debris disks to interpret the observed upper limits and to help plan future observations.

Detailed chemo-physical disk models \citep{Kamp:2004,Jonkheid:2004,Nomura:2005} have shown that the chemical structure of the surfaces of UV dominated protoplanetary disks resembles that of photon dominated regions seen at the surfaces of molecular clouds. The surface layers are dominated by far ultraviolet photons (6 - 13.6~eV) and contain mostly atomic hydrogen, and other atomic and ionized species. The transition to molecular species and thus to H$_2$ occurs roughly around a UV continuum optical depth $\tau({\rm 1000~\AA}) \sim 1$. This transition occurs very close to the disk surface and thus  the mass in this surface layer will naturally be small compared to the total disk mass. At very high gas temperatures (few 1000~K) however, the H/H$_2$ balance is dominated by H$_2$ destruction via collisions with neutral oxygen.

In the case of young active stars with high X-ray luminosity ($\sim 10^{28.5}-10^{31.5}$~erg~s$^{-1}$, \citealp{Glassgold:1997}), the chemistry and thus the depth of the surface layer is dominated by X-rays. Close to the star, X-ray heating dominates the disk energy balance, thus driving the chemistry that sets the H/H$_2$ balance.  In the outer disk, secondary ionization from X-rays determines the optical depth at which the transition from atomic to molecular hydrogen occurs.

Before we go into the details of the complicated chemo-physical disk models in the next section, we outline the basic physics governing the H/H$_2$ balance and hence the H\,{\sc i} mass fraction in the various evolutionary stages from young protoplanetary to very old debris disks.

\subsection{Flaring protoplanetary disks}

\begin{figure}[h]
\begin{center}
\includegraphics[width=16cm]{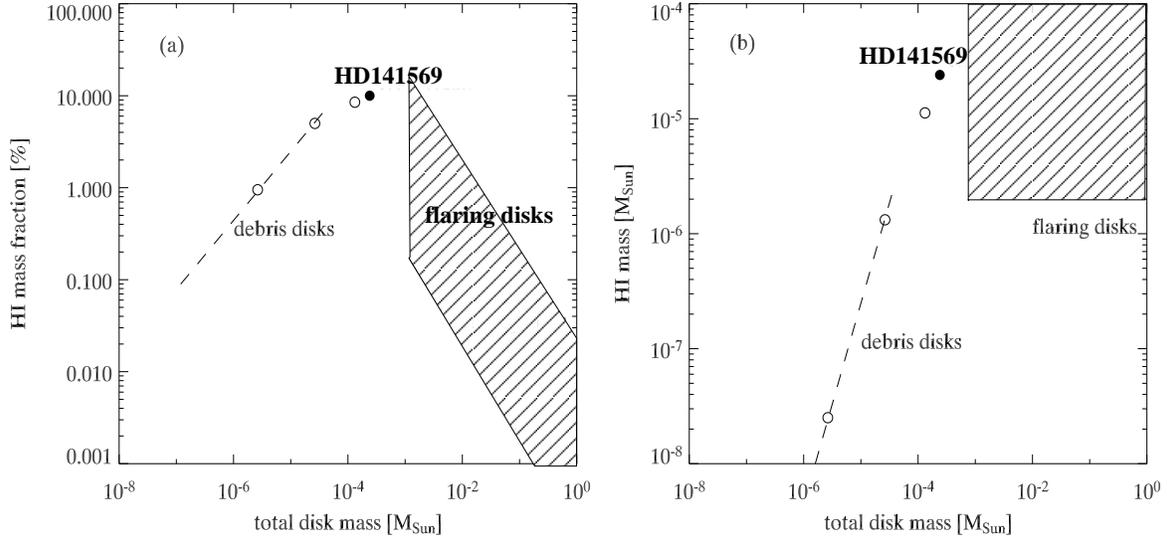}
\caption{(a) Mass fraction of neutral hydrogen as a function of total disk mass. Flaring disks are shown as solid lines, whereas dashed lines are used for debris disks. The filled circle denotes the H\,{\sc i} mass fraction in the HD\,141569A disk model and the open circles denote the respective values for the series of $\beta$~Pictoris models. Note that the $\beta$~Pictoris models all have the same dust mass, the gas-to-dust mass ratio therefore differs for different models. The shaded area illustrates the range of atomic hydrogen fraction in the presence of X-rays. (b) Same for total H\,{\sc i} mass.}
\label{HImassfrac}
\end{center}
\end{figure}

Where X-rays can be neglected, the hydrogen chemistry is driven by UV irradiation.
We can estimate the H\,{\sc i} mass in the disk by assuming that the optically thin surface layer of the disk contains only atomic hydrogen and that the transition to molecular hydrogen occurs around $\tau_{\rm in}({\rm 1000~\AA}) \sim 1$. This might slightly overestimate the mass of atomic hydrogen as we do not take into account H$_2$ self-shielding. However, we do take into account the flaring angle of the incoming radiation. Given the UV absorption cross section of our disk models, $\sigma({\rm 1000~\AA}) = \sigma({\rm UV}) = 5.86\times 10^{-22}$~cm$^{-2}$~(H-atom)$^{-1}$ and a typical flaring angle of $\alpha = 0.05$ (appropriate for the typical luminosity of T Tauri stars, $L_\ast = 1.4$~L$_\odot$), we obtain a total column density of
\begin{equation}
N_{\rm tot} = \frac{\alpha}{\sigma} = 8.5\times10^{19}~{\rm cm^{-2}}\,\,\,
\end{equation}
for the optically thin surface layer. Assuming that the disk extends from 5 to 300~AU, this translates into a total mass of $2\times 10^{-6}$~M$_\odot$. Adding mass to an optically thick disk simply adds mass to the interior, while the surface layer remains mostly unchanged. This makes the H/H$_2$ fraction of the disk diminish with increasing disk mass  (Fig.~\ref{HImassfrac}).  Herbig Ae stars are generally more luminous and hence their typical flaring angle is up to $50$~\% larger than that of the T Tauri stars. This increases the H/H$_2$ mass fraction by the same amount. 

The pure UV case presents a lower limit to the estimated H\,{\sc i} mass in the disk. In the presence of X-rays, the column density of atomic hydrogen in the disk surface increases substantially. \citet{Glassgold:2006} assumed a typical X-ray luminosity for T Tauri stars of $2\times 10^{30}$~erg~s$^{-1}$. They find a total column density of $N_{\rm tot} \sim 8.5\times10^{21}$~cm$^{-2}$ for the H\,{\sc i} surface. This number is fairly constant over the 5-20~AU range and hence we use it as an upper limit for the H\,{\sc i} mass increase in the presence of X-rays. Such X-ray luminosities may lead to a significantly larger H/H$_2$ mass fractions compared to the pure UV case. The shaded area in Fig.~\ref{HImassfrac} indicates thus the possible range of atomic hydrogen masses ranging from the pure UV to the UV plus X-ray case.
\clearpage
\clearpage
\subsection{Transition phase disks}

Massive protoplanetary disks are believed to evolve through a transition phase into debris disks. These transition phase disks are usually characterized by an absence of near-IR excess and thus large inner holes. The outer disks however, are still optically thick. Observations of the very few known transition phase disks reveal that they generally have a very complex structure (e.g. rings, spirals, clumps \citealp{Mouillet:2001,Grady:2001,Fukagawa:2004}), which is difficult to treat analytically. 

We will use later in this paper a disk model for the transition phase object HD\,141569A \citep{Jonkheid:2006} to fill the gap between the optically thick young disks and the optically thin debris disks. This model has been constructed to fit the dust scattered light and CO line observations of HD\,141569A. 

An UV only model predicts a relatively large H\,{\sc i} flux. In fact, Fig.~\ref{HImassfrac} shows that the H\,{\sc i} mass fraction peaks around the regime of transitional objects such as HD\,141569A. This is partly due to the fact that such transition phase disks are more tenuous than massive disks and hence UV photons penetrate deeper.

\subsection{Debris disks}
\label{massfrac:debris}

In very low mass disks, the H\,{\sc i} mass fraction becomes 100\% as the disk becomes totally optically thin even to UV line photons. However, such disks are very unrealistic, because generally dust grain growth and gas dispersal go along with a decreasing disk mass. Hence, there is a natural transition from protoplanetary disks to debris disks.

In the next section, we compute debris disk models which apply to the UV only case and the predicted H\,{\sc i} presents a lower limit for very active stars such as AU Mic, which have a significant X-ray luminosity. These models use gas-to-dust mass ratios varying between 100 (i.e. value typical of molecular clouds) and 2. This affects the formation of molecular hydrogen as the increasing dust surface area per gas changes the gas thermal balance and hence the temperature dependent H$_2$ formation rate. The H$_2$ formation time scale is generally faster than 1000~yr even in the very low mass debris disks and equilibrium chemistry is thus a valid assumption. 

In debris disks, we generally have a size distribution of dust grains with a minimum and maximum grain size, $a_{\rm min}$ and $a_{\rm max}$. However, the H$_2$ formation depends to first order only on the grain surface per hydrogen atom $n_{\rm d} \sigma_{\rm d}$. A distribution of grain sizes, can hence to first order be replaced by grains with a fixed size $a$, where $a$ is given by $a^2 = a_{\rm min} a_{\rm max}$ \citep[p.276]{Kruegel:2003}. In the following, we have chosen a fixed grain size $a = 3~\mu$m.

Since the dust surface area per hydrogen atom $n_{\rm d} \sigma_{\rm d}$ stays constant in our models, the main impact of a decreasing gas-to-dust mass ratio is an increasing gas temperature and thus larger thermal velocities $v$. This increases the collision rate between hydrogen atoms and grains and hence the H$_2$ formation rate. The photodissociation of H$_2$ is not affected by the gas-to-dust mass ratio. So, the H/H$_2$ balance shifts towards molecular hydrogen as the gas-to-dust ratio decreases. We use the detailed results from our models described in Sect.~\ref{disk models} to approximate the H\,{\sc i} mass fraction as a function of the gas-to-dust mass ratio $\delta$. Fig.~\ref{HImassfrac} shows that the H\,{\sc i} mass fraction scales approximately as $\delta^{0.7}$ (dashed line) for the debris disks.

\section{Detailed Disk Models}
\label{disk models}

\begin{figure}[h]
\begin{center}
\includegraphics[width=15cm]{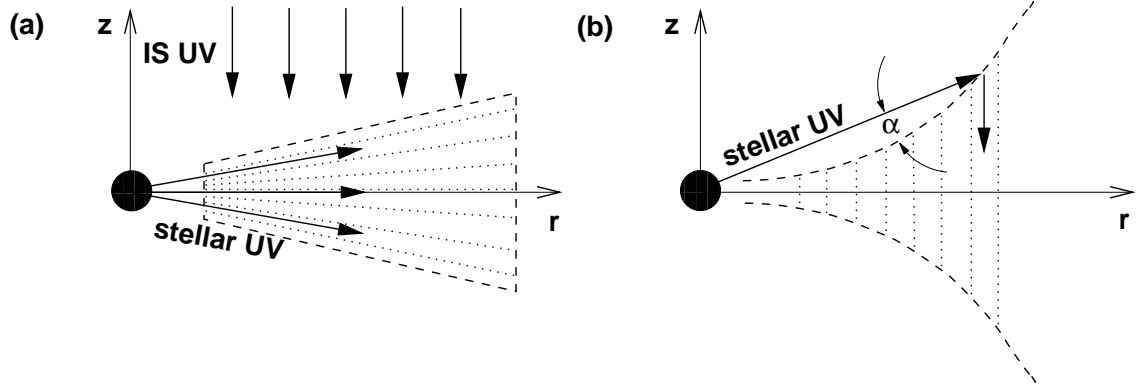}
\caption{Geometries and illumination for the (a) debris disks and (b) flaring disks. The dotted lines indicate how the models discussed in Sect.~\ref{disk models} are split up in slices. The debris disks are calculated from top to midplane following radial cuts, the flaring disks from top to midplane following vertical cuts.}
\label{sketch}
\end{center}
\end{figure}

This section describes the numerical models which we construct for the circumstellar disks. In a later section we use these models to predict the H\,{\sc i}~21~cm line emission.
We use two different types of disk models here for the optically thin and thick systems. Our models do not include X-rays and deal with the pure UV case. The debris disks are typically optically thin and thus we solve the chemistry and heating/cooling along radial rays starting in the top disk layers and moving down to the midplane. In the case of optically thick models, we divide the disk in vertical slices and calculate the coupled chemistry and heating/cooling in each slice. The different geometries are shown schematically in Fig.~\ref{sketch}. These two different approaches are necessary, because in the debris disk systems, the main source of energy input is the direct stellar irradiation, while in the younger optically thick  flaring systems, the irradiation is scattered from the surface into the disk. In the following, we summarize these two disk modeling approaches and the free parameters entering the disk models.

\clearpage
\clearpage

\subsection{Disk physics and chemistry of optically thin disks}
\label{disk models thin}

The basic physics and chemistry of these models is described in \citet{Kamp:2000}, \citet{Kamp:2001}. Here we only briefly summarize the salient features of these models. The disk density structure is assumed to be, $n(r,z) \approx r^{-p} \exp{-z^2/2 h^2}$, where we further assume that the scaleheight $h$ scales linearly with radius, and $h/r = 0.15$. The radial power law exponent $p$ is set to 2.5. The chemistry is modelled using a network consisting of 48 different species covering the elements H, He, C, O, S, Mg, Si, and Fe. We use typical IS abundances for these elements (Table~\ref{abus}). These species are connected through 281 reactions, including also cosmic ray chemistry, photochemistry and the chemistry of excited H$_2$. The dust temperature is derived from the assumption of large black body grains in radiative equilibrium around the star, $a = 3~\mu$m, $\rho = 3.0$~g~cm$^{-3}$. The gas temperature is derived from a detailed energy balance including the most relevant heating and cooling processes \citep{Kamp:2001}. Table~\ref{diskmodel_param} lists the basic parameters of the computed disk models.  
\clearpage
\begin{table}[htdp]
\caption{Element abundances in the disk models normalized to a hydrogen density of $10^{12}$~cm$^{-3}$ for debris disk models}
\begin{center}
\begin{tabular}{lr|lr}
element & abundance \,\,\,\,\,& element & abundance \\[2mm]
\hline
H     &  12.00  \,\,\,\,\,& S     & 6.90  \,\,\,\,\,\\
He   &  11.00  \,\,\,\,\,& Mg  & 6.00  \,\,\,\,\,\\
C     &    8.15  \,\,\,\,\,& Si    &  5.90  \,\,\,\,\,\\
O     &    8.51  \,\,\,\,\,& Fe   &  5.40  \,\,\,\,\,\\
\end{tabular}
\end{center}
\label{abus}
\end{table}%
\clearpage
H$_2$ photodissociation occurs in the wavelength range between 912 and 1110~\AA . We calculate the photodissociation rate from
\begin{equation}
\Gamma = 4.2 \times 10^{-11} ( \chi f_{\rm shield,r} + G_0 f_{\rm shield,v} ) n({\rm H_2})\,\,\,{\rm s^{-1}~cm^{-3}}\,\,\,,
\end{equation}
where $\chi$ is the integrated stellar radiation field between 912 and 1110~\AA\  normalized to the Habing field \citep{Habing:1968}. The stellar radiation field is an ATLAS9 model \citep{Kurucz:1992} with stellar effective temperatures as listed in Tab.~\ref{diskmodel_param}. An interstellar radiation field with $G_0=1$ in units of the Habing field is added to the stellar one. The H$_2$ self shielding is approximated by Eq.(37) of \citet{Draine:1996} for radial ($f_{\rm shield,r}$) and vertical directions ($f_{\rm shield,v}$). We assume here that radiative decay of vibrationally excited H$_2$ is faster than direct photodissociation out of these levels. This approximation is valid for $\chi < 10^5$ \citep{Shull:1978}, a limit that holds for all models, except the inner 5-10~AU of the T Tauri disk and the inner 1.4-2~AU of the Herbig Ae disk. Even though the H/H$_2$ fraction in those areas might change if direct photodissociation from vibrationally excited levels is included, the mass of those inner disks is negligible compared to the total disk mass. The predicted H\,{\sc i} line profiles and fluxes will hence not be significantly affected.

The H$_2$ formation rate is
\begin{equation}
R_{\rm form} = 0.5 n_{\rm H} n_{\rm d} \sigma_{\rm d} v \epsilon_{\rm H_2} S(T) ~~{\rm s^{-1} cm^{-3}}\,\,\, ,
\end{equation}
where $n_{\rm H}$ is the hydrogen number density, $n_{\rm d}$ is the dust grain number density, $\sigma_{\rm d}$ is the dust grain cross section, $v$ the thermal velocity, $\epsilon_{\rm H_2}$ the recombination efficiency of hydrogen atoms on the surface and $S(T)$ a temperature dependent sticking coefficient for atomic hydrogen on the grain surface. Higher gas temperatures lead to larger thermal velocities and hence to more H-grain collisions. The recombination efficiency depends on the  details of the H$_2$ formation mechanism. \citet{Cazaux:2002} approximate the recombination efficiency $\epsilon_{\rm H_2}$ as a function of grain temperature. While at low grain temperatures nearly every incoming H-atom recombines with a surface H-atom to form H$_2$, this process becomes much less efficient at higher temperatures (50-100~K), where direct re-evaporation of pysisorbed H-atoms competes with H$_2$ formation. The sticking probability $S(T)$ can in principle depend on temperature, but here we assume it to be equal to one.

\begin{figure}[h]
\begin{center}
\includegraphics[width=9cm]{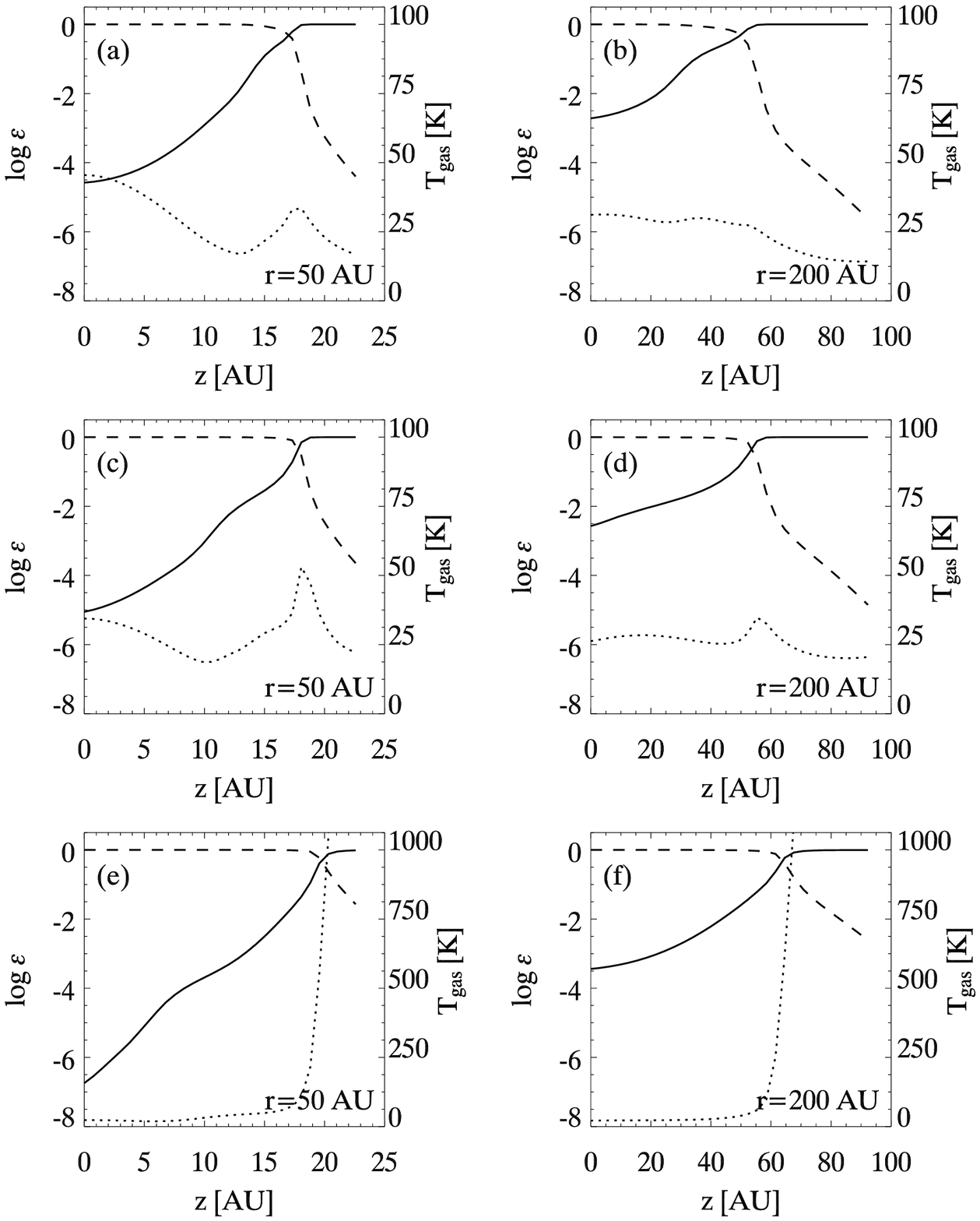}
\caption{H/H$_2$ transition for two vertical cuts at 50 and 200~AU in the three debris disk models: (a), (b) 44~M$_{\rm Earth}$, (c), (d) 8.8~M$_{\rm Earth}$, (e), (f) 0.88~M$_{\rm Earth}$. The solid lines are the H abundance, the dashed lines the H$_2$ abundance. The dotted lines denote the gas temperature using the scale on the right axis.}
\label{hh2_debris}
\end{center}
\end{figure}

The models for $\beta$~Pictoris are calculated assuming a constant dust mass of $1.3\,10^{-6}$~M$_\odot$ derived from fitting the infrared spectral energy distribution \citep{Chini:1991}. Since the total gas mass in the disk around this star is not yet well constrained, we assume here different gas-to-dust mass ratios ranging from the canonical value of 100 down to 2. Fig.~\ref{hh2_debris} shows the H and H$_2$ abundances for two vertical cuts at 50 and 200~AU in the three disk models. This quantitatively illustrates what has been stated in Sect.~\ref{massfrac:debris}: the gas temperature increases as the gas-to-dust mass ratio decreases. The heating/cooling balance shifts, because the line cooling decreases with decreasing gas densities, while the photoelectric heating stays constant.
\clearpage
\clearpage
\begin{table}[htdp]
\caption{Parameters of disk models:  debris disk models, HD\,141569A is a transition phase disk model and flaring disk models for a Herbig Ae and T Tauri star}
\begin{center}
{\footnotesize
\begin{tabular}{lllllllllll}
   type                &  $T_{\rm eff}$ &    R$_\ast$      &   M$_\ast$       & M$_{\rm disk}$   & M$_{\rm gas}$/M$_{\rm dust}$ & R$_{\rm in}$  & R$_{\rm out}$ & a & $\sigma$(UV) & L$_{\rm UV}$\\
                             &    [K]                 &    [R$_\odot$] &   [M$_\odot$]  &  [M$_\odot$]         && [AU]                 &  [AU]  & [$\mu$m] & [cm$^2$/(H-atom)] &[erg s$^{-1}$] \\[2mm]
\hline\\[-2mm]
debris disk & 8250 & 1.7 & 2.0 & $1.3\times10^{-4}$      & 100 & 40                    & 500 & 3   & $2.34\times 10^{-23}$ & $1.9\times 10^{27}$\\
                             & & & & $2.6\times10^{-5}$      &   20 &  40                    & 500 & 3  & $1.17\times 10^{-22}$ & $1.9\times 10^{27}$\\
                             & & & & $2.6\times10^{-6}$      &     2 &  40                    &  500 & 3  & $1.17\times 10^{-21}$ & $1.9\times 10^{27}$\\[3mm]
HD\,141569A  & 10\,000 & 1.7 & 2.0 & $2.4\times10^{-4}$      & 100 & 80                    & 500 & 0.1& $5.86\times 10^{-22}$ & $2.7\times 10^{33}$ \\[3mm]
Herbig Ae       & 9500 & 2.4 & 2.5 & 0.01                       & 100 & 1.4                   & 300 & 0.1 & $5.86\times 10^{-22}$ & $6.6\times 10^{29}$\\
T Tauri             & 4000 & 2.5 & 0.5 & 0.01                        & 100 &  5.0                  & 300 & 0.1 & $5.86\times 10^{-22}$ & $2.0\times 10^{31}$\\
\end{tabular}
}
\end{center}
\label{diskmodel_param}
\end{table}
\clearpage

\subsection{Disk physics and chemistry of optically thick disks}
\label{disk models thick}
 
For the massive optically thick disk models we follow a similar approach as for the optically thin models. However, the underlying equilibrium density structure is very different; it is taken from two-dimensional hydrostatic equilibrium models  \citep{Dullemond:2002}. In these models, the gas temperature is first approximated by the dust temperature found from detailed continuum radiative transfer. We then re-evaluate the gas temperature in the context of the coupled chemistry and heating/cooling balance \citep{Kamp:2004}. Another difference from the hydrostatic equilibrium models is need to account for the flaring angle, that is the grazing angle under which the stellar irradiation hits the disk surface (see Fig.~\ref{sketch}).  

\begin{figure}[htbp]
\begin{center}
\includegraphics[width=10cm]{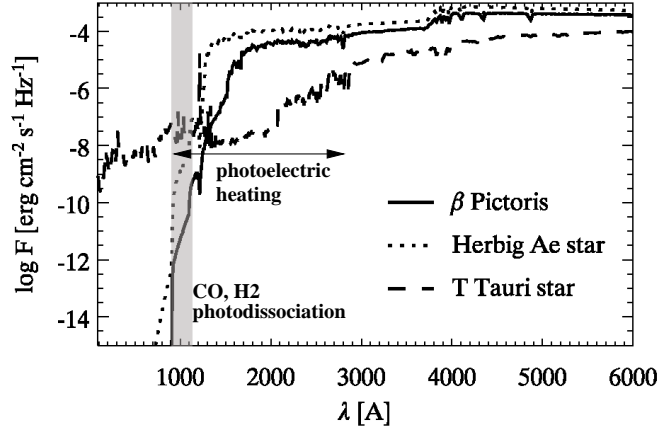}
\caption{UV radiation fields for the various disk models: ATLAS9 $\beta$~Pictoris model (solid line), 
ATLAS9 Herbig Ae star model (dotted line), T Tauri model with UV excess (dashed line).}
\label{UVrads}
\end{center}
\end{figure}

The stellar radiation field for a T Tauri star is a combination of an ATLAS9 model and a scaled solar chromosphere model as outlined in \citet{Kamp:2004}. Even though the chromospheric UV excess contributes little to the total luminosity and hence the flaring of the disk, it is relevant for the photodissociation of molecules and the photoelectric heating of the disk. In the case of the Herbig Ae star, we used a pure ATLAS9 photospheric model with $T_{\rm eff}=9500$~K and $\log g=4.0$. Fig.~\ref{UVrads} compares the various UV and optical radiation fields used in the disk models. Diffuse interstellar UV irradiation is not relevant for young disks at radii smaller than 300~AU.

The chemical network and heating/cooling processes are the same as for optically thin disks. The dust properties however are different reflecting a much younger evolutionary stage. We use the same grain properties and opacities as used in the underlying continuum radiative transfer models by \citet{Dullemond:2002} , $a=0.1$~$\mu$m, $\sigma({\rm UV})=5.86 \times 10^{-22}$~cm$^2$/H-atom. The  protoplanetary disk models described in Table~\ref{diskmodel_param} have been previously published in \citet{Kamp:2004} and \citet{Kamp:2006} as a typical T Tauri and Herbig Ae star respectively.   

\section{H\,{\sc i}~21~cm line radiative transfer}
\label{radtrans}

We assume here that the level population numbers are in LTE, i.e.\ T$_{\rm ex}=$ T$_{\rm gas}$, which is a valid approximation for n$_{\rm H} \gg 4\,10^{-4}\,T_{\rm gas}$. Given typical particle densities and gas temperatures of $10^3< n_{\rm H} < 10^{11}$~cm$^{-3}$ and 10--10\,000~K, this assumption is fullfilled throughout the entire disk. We thus assume a ratio of $n_1/n_0 = 3$ for the level population numbers, a good approximation for $T_{\rm gas} > 10$~K (error smaller than 1\%).  Even though the gas temperature does not play a role in the H\,{\sc i} level populations, it is still crucial for the determination of the H/H$_2$ chemical balance and hence the total number density of atomic hydrogen.

The line emission and absorption coefficients $\epsilon_{10}$ and $\kappa_{01}$ are derived from
\begin{eqnarray}
\epsilon_{10} &=& h \nu_{10}\,A_{10}\, n_1({\rm H})~~~{\rm erg}~{\rm s}^{-1}~{\rm cm}^{-3} \\
\kappa_{01}   &=& \frac{h \nu_{10}}{c}\,B_{01}\,n_0({\rm H})\,\frac{h\nu_{10}}{kT} ~~~{\rm s}^{-1}~{\rm cm}^{-1}
\end{eqnarray}
The absorption coefficient is corrected for stimulated emission assuming LTE level population numbers and using the first order term of the expansion of the exponential.
The basic atomic parameters for the line are $A_{10} = 2.87\,10^{-15}$~s$^{-1}$, $\nu_{10} = 1.4205$~GHz, and $n_1({\rm H}) \sim 3/4\times n({\rm H})$. The total emission per velocity bin $dv$ is derived from an integration along the optical path within the beam area $dA$ and within the respective velocity bin $dv=$[$v_i, v_{i+1}$]
\begin{equation}
F_{10} = \int_{\rm beam} \frac{\epsilon_{10}}{dv} \, \exp(-d\tau_{10}) \,ds~~~{\rm erg}~{\rm s}^{-1}~{\rm km~s}^{-1} \,\,\, .
\end{equation}
The optical depth $d\tau_{10}$ is derived from the absorption coefficient
\begin{equation}
d\tau_{10} = \int \frac{\kappa_{01}}{dv}\,ds \,\,\,.
\end{equation}
The typical 'beam averaged' optical depth are smaller than $\tau \sim 0.001$ for the Herbig Ae disk seen edge-on and $\sim 0.02$ for the pole-on model.  Even though the column densities are smaller in the pole-on model, the velocities in the line of sight are zero; hence the optical depth contribution at $v=0$ is maximized.

The optical depth is -- in the case of a rotating disk -- defined as a function of velocity, $\tau(v)$. Hence, the corresponding column density to derive the optical depth is the averaged column density of all H-atoms falling in the velocity bin $dv$ averaged over the beam. We assume that the disk is in keplerian rotation $v_{\rm kep}$ around the star with central mass $M_\ast$. The velocity profile for an inclined disk (the inclination angle $i$ is defined between the line of sight and the disk midplane, so that $i=0$ denotes an edge-on disk ) is given by
\begin{equation}
v_{\rm rot}(r,z,\phi) = v_{\rm kep}(r,z) \sin(\phi) \cos(i)\,\,\, ,
\end{equation}
where $\phi$ is the azimutal angle in the disk.

To calculate observed 21 cm line fluxes, the disk is projected onto a Cartesian grid with the z-axis in the direction of the line of sight. A separate cube is created for each velocity bin (see Fig.5). For each of the cubes, the emission and optical depth are integrated along the line of sight.

\clearpage
\begin{figure}[h]
\begin{center}
\includegraphics[width=9cm]{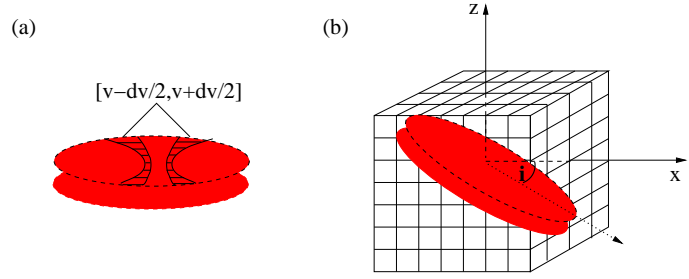}
\caption{Velocity binning (a) and coordinate transformation (b) for the 21~cm line radiative transfer.
The shaded area in (a) corresponds to the velocity bin  [$v-dv/2,v+dv/2$] and matter within that area will contribute to the line emission profile at the velocity $v$.}
\label{sketch_rad}
\end{center}
\end{figure}
\clearpage

The intensity arriving at Earth is given by
\begin{equation}
I_{10} = \frac{1}{4\pi d^2} \int F_{10} dA~~~{\rm erg}~{\rm cm}^{-2}~{\rm s}^{-1}~{\rm km~s}^{-1} \,\,\, .
\end{equation}
Converting to more common units of erg cm$^{-2}$ s$^{-1}$ Hz$^{-1}$ is done by using
 \begin{equation}
dv = c \frac{d\nu}{\nu_{10}} \,\,\, .
\end{equation}
We use the beamwidth given in Table~\ref{upperlimits} as a FWHM to convolve the resulting images in each velocity channel assuming a gaussian beam profile $g(x,y)$. This leads to the final intensity in observers units, mJy per beam, extracted at the center $(x_1,y_1)$
\begin{equation}
I_{10} (x_1,y_1) = 10^{26} \int g(x,y) I_{10}(x_1-x,y_1-y) dx dy ~~~{\rm mJy / beam}\,\,\,.
\end{equation}

\section{Predictions for H\,{\sc I} disk emission}
\label{predictions}

 We will constrain disk parameters, especially the total disk gas mass, by comparing the observed upper limits for the H\,{\sc i}~21~cm line presented in Sect.~\ref{observations} to predicted emission from the individual disk models. In Sect.~\ref{hfrac} we presented rough estimates of the H\,{\sc i} mass in the various disks based on various simplifications. In this section we take a more quantitative approach involving detailed disk models as well as stellar parameters, inclinations and distances appropriate for the individual sources. We apply the models described in Sect.~\ref{disk models} to predict H\,{\sc i} line maps and intensities based on the observed beam sizes. Table~\ref{diskmodel_results} summarizes the basic H\,{\sc i} characteristics of those disk models.
\clearpage
\begin{table}[htdp]
\caption{H\,{\sc i}~21~cm line characteristics of the disk models}
\begin{center}
\begin{tabular}{llllll}
   type                &  M$_{\rm disk}$   & H\,{\sc i} mass         & beam size & peak intensity &  peak surface brightness\\
                             &  [M$_\odot$]         &  [M$_{\rm Earth}$]  &       [AU]       & [mJy]               &  [$\mu$Jy/arcsec$^2$]\\[2mm]
\hline\\[-2mm]
debris disk & $1.3\times10^{-4}$	&  4.0 			& 1280 & 18.0 & 480 \\
                             & $2.6\times10^{-5}$	&   0.4		         	& 1280 &   1.8  & 40 \\
                             & $2.6\times10^{-6}$	&  $9\times10^{-3}$ 	& 1280 &  $4.4\times 10^{-2}$ & 1\\[3mm]
HD\,141569A    & $2.4\times10^{-4}$	& 13.0			& 1100 & 1.6     & 290 \\[3mm]
Herbig Ae       & 0.01                       	     	&  0.3			& 980   & 0.05   &110 \\
T Tauri             & 0.01                          	&  1.3			& 1120 & 0.24    & 400\\
\end{tabular}
\end{center}
\label{diskmodel_results}
\end{table}%
\clearpage
\subsection{$\beta$~Pictoris}

The disk around $\beta$~Pictoris is the best studied example of a debris disk. O\,{\sc vi} line emission has been detected with FUSE \citep{Deleuil:2001} indicating some level of stellar activity. A followup search for the O\,{\sc vii} line with XMM-Newton has revealed a very low flux of $6\times10^{-16}$~erg~cm$^{-2}$~s$^{-1}$ \citep{Hempel:2005}. This results in a X-ray to UV luminosity ratio of $10^{-2}$. The impact of such an X-ray flux on the chemistry of debris disks (gas-to-dust mass ratios smaller than 1) is yet unknown. We can therefore not rule out the possibility that the disk around $\beta$~Pictoris contains additional atomic hydrogen produced by X-rays.

\begin{figure}[htbp]
\begin{center}
\includegraphics[width=18cm]{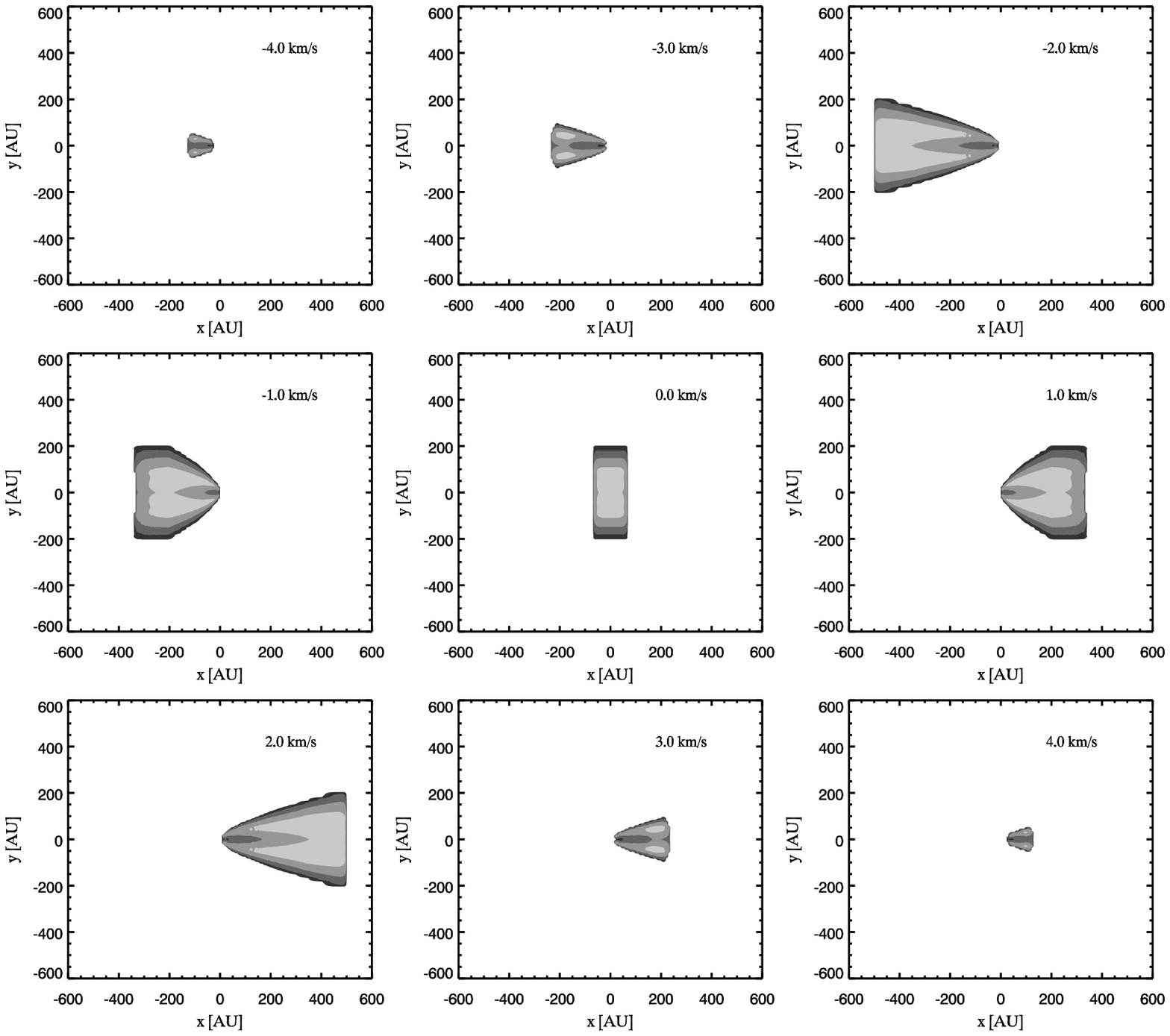}
\caption{Map of the atomic hydrogen distribution in the $\beta$~Pictoris 44~M$_{\rm Earth}$ disk model (gas-to-dust mass ratio of 100,  inclination 0$^\circ$). The different gray shades denote steps of a factor 10 in emission.}
\label{betapic:map}
\end{center}
\end{figure}

\begin{figure}[htbp]
\begin{center}
\includegraphics[width=8cm]{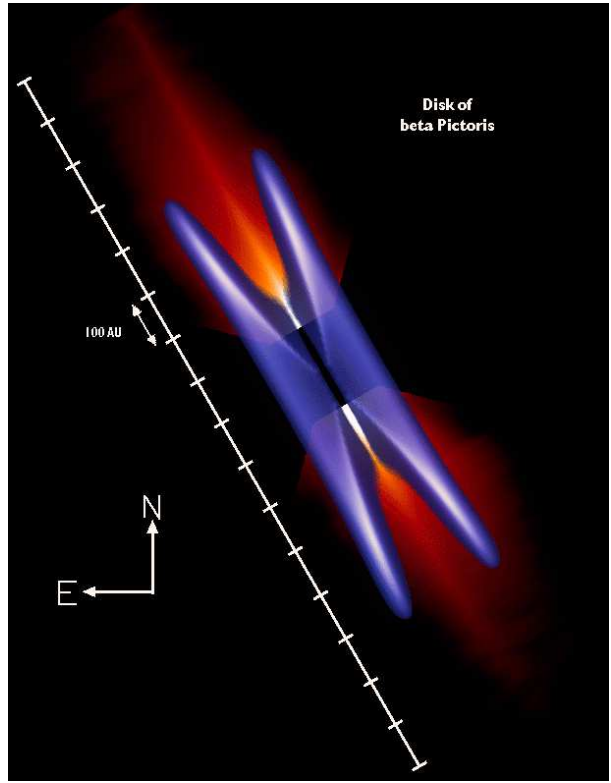}
\caption{H\,{\sc i}~21~cm line image from the 4.4~M$_{\rm Earth}$ mass model (blue) on top of the HST optical scattered light image taken with Wide Field Planetary Camera 2 (red, credit Al Schultz and his team). The maximum H\,{\sc i} emission (white) comes from the surface of the gaseous disk; peak fluxes are $17 \mu$Jy/arcsec$^2$, well within the reach of the SKA.}
\label{posterbetapic}
\end{center}
\end{figure}

The dust mass in the $\beta$~Pictoris disk is well known from previous observations and we assume in this paper a fixed dust mass of 0.44 M$_{\rm Earth}$ (see Sect.~\ref{disk models thin}). We computed models with gas-to-dust mass ratios from 100 to 2. Fig.~\ref{betapic:map} shows the distribution of atomic hydrogen in the disk for individual velocity channels ($\Delta v = 1.0$~km/s). The atomic hydrogen is located in a thin layer on top of the entire disk. The disk towards the midplane looks filled by atomic hydrogen, but this is due to the projection of the three dimensional disk model into the plane of observations. It stems from H\,{\sc i} in the disk surface outside of the observing plane. This can be seen in Fig.~\ref{posterbetapic}, where we plot the integrated H\,{\sc i}~21~cm line image on top of the WFPC2 dust scattererd light image of Al Schultz and his team (private communication). As expected, the maximum of the H\,{\sc i} emission occurs at the surface of the gaseous disk. Note also that the HI emission lies well within the region from which scattered light from dust grains can be detected. Recent imaging of the disk in the Na\,{\sc i} D lines indicates scaleheights of $h/r \sim 0.3$ \citep{Brandeker:2004}, a factor 2 higher than what our models actually use. It is currently unclear which processes lead to such large scaleheights, but one reason might be higher gas temperatures in case of a very low gas-to-dust mass ratio. As explained in Sect.~\ref{disk models thin}, this is due to a shift in the thermal balance between photoelectric effect and fine structure line cooling. The distribution of the micron sized dust grains responsible for the scattered light is sustained by collisions among dust grains and/or larger bodies within the disk and is not coupled to the gas.

\begin{figure}[htbp]
\begin{center}
\includegraphics[width=8cm]{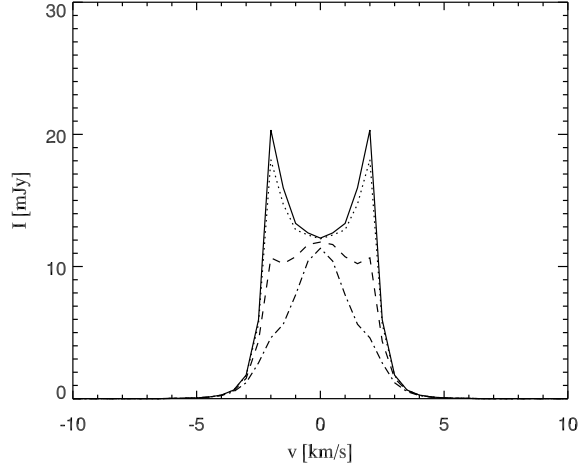}
\caption{{21 cm line intensity for the $\beta$~Pictoris 44~M$_{\rm Earth}$ disk model (gas-to-dust mass ratio of 100,  inclination 0$^\circ$) and four different beam sizes: 72.5" (1400~AU, solid line), 57" (1100~AU, dotted line), 33.7" (650~AU, dashed line), and 22.3" (430~AU, dash-dotted line).}}
\label{betapic:beam}
\end{center}
\end{figure}

\begin{figure}[htbp]
\begin{center}
\includegraphics[width=8cm]{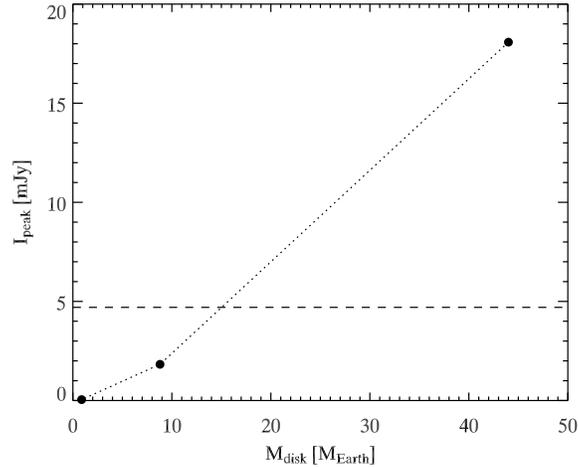}
\caption{21 cm line peak intensities for different gas-to-dust ratios in the $\beta$~Pictoris model: 100 (M$_{\rm gas} = 44$~M$_{\rm Earth}$), 20 (M$_{\rm gas} = 8.8$~M$_{\rm Earth}$), and 2 (M$_{\rm gas} = 0.88$~M$_{\rm Earth}$). Line fluxes were calculated with a gaussian beam with FWHM of 66.5" (1280 AU). The dashed line denotes the upper limit derived from the ATCA observations (Table~\ref{upperlimits}) for this beamsize.}
\label{betapic:mass}
\end{center}
\end{figure}

We checked the resulting line intensities for the different beam sizes used in the observations. The beams were always centered on the stellar position. It turns out that the 66.5" (1280~AU) beam includes most of the H\,{\sc i} emission of $\beta$~Pictoris, which is not surprising given our disk model extension of $r=500$~AU (see Fig.~\ref{betapic:beam}). The smaller beams miss a large fraction of the emission at intermediate velocities ($v \sim 2$~km/s), which is located in the outer regions of the disk (Fig.~\ref{betapic:map}).  Given the density law of the underlying disk model, most of the disk mass --- and hence also the H\,{\sc i} mass ---  resides in the outer parts of the disk. The maximum H\,{\sc i} intensity of the 44~M$_{\rm Earth}$ disk model, 18~mJy, is larger than the observed upper limit of 4.7~mJy. The ATCA observations (F95) therefore imply an upper limit of $\sim15$~M$_{\rm Earth}$ on the total gas mass (H\,{\sc i}+H$_2$) in the disk around $\beta$~Pictoris (Fig.~\ref{betapic:mass}). 

\clearpage

\clearpage
\subsection{LkCa15}

For modeling the H\,{\sc i} emission from this star, we used the T Tauri disk model by \citet{Kamp:2004}. This model extends from 5 to 300 AU and has a mass of 0.01~M$_\odot$. The distance to the star is 140~pc and \citet{Qi:2003} have shown from modeling CO and HCO$^+$ that the inclination is $57^\circ  \pm 5^\circ$. LkCa 15 has not been detected in the ROSAT All Sky Survey (RASS). \citet{Neuhaeuser:1995} put an upper limit of $4\times10^{29}$~erg~s$^{-1}$ on the X-ray luminosity for this source. The resulting upper limit to the X-ray to UV luminosity ratio is 0.02.

The H\,{\sc i} observations with a beam of 8$^"$ comprised the entire disk as opposed to the smaller beam of 2.5$^"$, which contains only the inner 125~AU of the disk. Using the larger beam, we predict a 21~cm line flux of 0.24~mJy, two orders of magnitude lower than the upper limit derived from GMRT observations. For the smaller beam size, we obtain fluxes of 0.13~mJy. The observed upper limit of 20~mJy to the 21cm flux corresponds to an upper limit of $3 \times 10^{-4}$~M$_\odot$ for the atomic hydrogen mass. 

Since we have not accounted for the H\,{\sc i} production by
X-rays, in young active stars which are strong X-ray emitters, the H\,{\sc i} mass would be much higher than what we obtain. As can be seen in Fig.~\ref{HImassfrac} for such stars in nearby (50-100 pc) young associations, the 21~cm line flux might be detectable.

\subsection{HD\,163296}

For modeling the H\,{\sc i} emission from this star, we used the Herbig Ae disk model by \citet{Kamp:2006}. The stellar effective temperature in this model of 9750~K is  only marginally higher than the value quoted in Table~\ref{diskmodel_param} and the stellar mass is 2.5~M$_\odot$. HD\,163296 is a weak X-ray source with $L_{\rm X} = 4\times 10^{29}$~erg~s$^{-1}$ \citep{Swartz:2005}, corresponding to an X-ray to UV luminosity ratio of 0.6. Therefore, the predicted H\,{\sc i} mass from our model should be considered as a lower limit.

The model extends from 1.7 to 300~AU and has a mass of 0.01~M$_\odot$. We use a distance of 122~pc and an inclination angle of 65$^\circ$ from \citet{Jaya:2001} and \citet{Dominik:2003} respectively. As in the case of LkCa15, we used the larger beam of 8$^"$, so that the beam contains the entire disk. The  peak H\,{\sc i}~21~cm flux for this model is 0.05~mJy. The fluxes are lower than for LkCa15, because the H\,{\sc i} fraction in the model is about an order of magnitude lower ($\sim 0.01$~\%). This is due to a weaker UV radiation field. While we assumed a chromospheric UV excess for the T Tauri star, we chose a pure photospheric model in the case of the Herbig Ae star; chromospheric activity is mostly restricted to solar-type stars. The observed upper limit of 13.7~mJy corresponds to an H\,{\sc i} mass upper limit of $3 \times 10^{-4}$~M$_\odot$.

\subsection{HD\,141569A}

Our estimates of atomic hydrogen fractions in disks suggested that the fraction would be maximal in transition phase objects. HD\,141569A which has been modeled extensively by \citet{Jonkheid:2006} is such a transition phase disk. This star has not yet been observed in the 21~cm line, but given that it is an excellent candidate for future observations, we present here predictions for the H\,{\sc i}~21~cm flux. 

The parameters for the dust phase of the model have been derived from scattered light observations of \citet{Mouillet:2001}, while the gas phase constraints come from a fit to the CO rotational emission (J=3-2) from \citet{Dent:2005}.  The dust is mostly confined in two rings located at 185 AU and 325 AU. {The gas follows} a smooth surface density profile $\Sigma \sim r^{-1.2}$ extending all the way in to 80 AU. This star has two M-type companions at distances of $790$ and $670$~AU that are most likely the source of the X-rays measured in the ROSAT All Sky Survey. The X-ray luminosity is $8.72\times 10^{29}$~erg~s$^{-1}$ \citep{Weinberger:2000}. Given the distance of the X-ray source from the disk around the primary star, the X-ray flux is a factor $5\times10^5$ lower than the UV flux at 40~AU. Thus,  X-rays will not affect the chemical processes around HD\,141569A.

The maximum emission is 1.6~mJy using a beam diameter of 11" (1100~AU). With the current instrument sensitivity, this would translate into a 1\,$\sigma$ signal. However, given the caveats of the above presented disk model, we cannot exclude that there might be more H\,{\sc i} than anticipated from CO observations. A different gas density profile or an underestimation of the UV radiation field of this star may lead to an even higher fraction of atomic hydrogen and hence to a larger signal. Clearly, this is the most promising candidate to look for atomic hydrogen.

\section{Discussion}
\label{discussion}

In the following, we summarize the main results of the H\,{\sc i} modeling and discuss them in the light of present and future observing possibilities.

\subsection{Gas in the disk around $\beta$~Pictoris}

The gas mass upper limit from the H\,{\sc i} observations is $\sim15$~M$_{\rm Earth}$. This amounts to a H\,{\sc i} midplane column density of $1.4\times10^{19}$~cm$^{-2}$.
However, the same model predicts H$_2$, O\,{\sc i}, C\,{\sc i}, and C\,{\sc ii} midplane column densities of $4.1\times10^{21}$, $1.8\times10^{18}$, $1.2\times10^{17}$, and $2.3\times10^{16}$~cm$^{-2}$. These values are much larger than those derived from UV absorption measurements by \citet{Lec:2001} and \citet{Roberge:2006}. The latter paper compares the elemental composition in the stable $\beta$~Pictoris gas to  the composition of the Sun, meteorites and comets; it leads to the conclusion that the gas in the disk around this star is extremely carbon overabundant, raising the interesting question of the origin of this peculiar chemical composition. 

Current H\,{\sc i}~21~cm line observations cannot rule out the presence of enough atomic hydrogen in the disk around $\beta$~Pictoris to obtain a solar gas composition - in terms of hydrogen versus metals;  metals here comprise all elements besides hydrogen and helium. More sensitive radio observations are needed to determine the true hydrogen content of this disk. Since current facilities such as the  GMRT and ATCA are limited to sensitivities of the order of 1-2 mJy, we have to wait for the Square Kilometer Array (SKA) to address this question \citep{Schilizzi:2004}. The SKA will have a roughly 10 times larger collecting area than current telescopes and a spatial resolution of up to 1~mas for imaging. The sensitivity of the SKA is expected to be at least a factor of $100$ above the one of current interferometers over a wide range of resolutions. 
        
Our models, however, suggest that the dominant form of hydrogen in debris disks is molecular and not atomic hydrogen (Sect.~\ref{massfrac:debris} and Sect.~\ref{disk models thin}). Thus the FUSE non-detection of H$_2$ in the $\beta$~Pictoris UV spectrum puts the most stringent upper limit on the disk hydrogen mass \citep{Lec:2001}. The H$_2$ column density from these observations is lower than $10^{18}$~cm$^{-2}$; this leads to an upper disk mass limit of $0.005$~M$_{\rm Earth}$ ($0.4$~M$_{\rm Moon}$), if we apply the same underlying disk structure  we used in our debris disk models.
 

\subsection{Future Observations using H\,{\sc i} as a tracer}

Our models show that the non-detection of the H\,{\sc i}~21~cm line is not due to the absence of atomic hydrogen in these disks, but rather due to the weakness of the signal as compared to the sensitivities of current telescopes. However, our models also show that the SKA will be able to detect H\,{\sc i}~21~cm emission from protoplanetary disks around stars in nearby star forming regions. These observations would yield a number of interesting constraints on the properties and evolution of these disks.

In massive disks, the H\,{\sc i}~21~cm line traces  the thin layer of the disk surface that is directly exposed to X-ray and UV irradiation; the H/H$_2$ mass fraction depends mainly on the strength of the radiation field impinging the disk surface. More stringent upper limits would hence give an independent check of the irradiation environment of the star. In addition, the velocity information in the line profile provides an independent measurement of the stellar mass, disk extension and evaporative flow from the disk surface. Typical velocities for the evaporative flow are larger than 20\% of the local sound speed \citep{Adams:2004}, 
\begin{equation}
c_s = \sqrt{G\,M_\ast/(0.5\,r)} = 42 \left(\frac{M_\ast}{M_\odot}\right)^{0.5} \left(\frac{r}{\rm AU}\right)^{-0.5}\,\,\,{\rm km~s}^{-1}\,\,\,.
\end{equation}
 Such an outflow should be detectable at the typical line resolutions of 1~km/s used in the radio regime.

At later stages, e.g.\ in the transition phase disk around HD\,141569A, the mass fraction of atomic hydrogen is of the order of 10\%. This makes H\,{\sc i} indeed a good probe of total disk mass in such transition phase disks. (See Fig.~\ref{HImassfrac}b for the  dependence of H\,{\sc i} mass on total disk mass). These disks are the best candidates for 21~cm line detection; our models indicate that existing telescopes may be able to detect emission from disks at distances smaller than 100~pc.

Debris disk systems often have a very low gas-to-dust mass ratio, which drives the hydrogen into its molecular form. In such debris disks, the H\,{\sc i} fraction is very low, smaller than 10\%.  The integrated H\,{\sc i} flux is therefore significantly lower than for the transitional phase disk. This scenario therefore predicts hugely different total H\,{\sc i} emission for the different types of disks which can in principle be tested observationally.

 Finally, we note that we have ignored H\,{\sc i} production by X-rays, which means that our models underestimate the H\,{\sc i} 21cm flux. Self consistent modeling of the X-ray chemistry as well as X-ray heating will hence probably lead to an even more optimistic prediction for the detectability of the 21cm emission from circumstellar disks by next generation radio telescopes.

\begin{acknowledgements}
The authors thank Massimo Robberto, David Hollenbach, Joan Najita and Jean-Charles Augereau for fruitful discussions during the preparation of this work. We acknowledge a constructive referee report that helped to improve the content and readability of the paper.
\end{acknowledgements}

\end{document}